\documentclass[aps,pra,twocolumn,groupedaddress,notitlepage,showpacs,floatfix,superscriptaddress]{revtex4-1}
\pdfoutput=1

\usepackage{graphicx,graphics,epsfig,subfigure,times,bm,bbm,amssymb,amsmath,amsthm,mathrsfs,MnSymbol}
\usepackage{gensymb}
\usepackage{amsfonts}
\usepackage[matrix,frame,arrow]{xypic}
\usepackage[pdfstartview=FitH]{hyperref}
\usepackage{eucal}
\hypersetup{
    colorlinks=true,       
    linkcolor=red,          
   citecolor=magenta,        
    filecolor=magenta,      
    urlcolor=cyan,           
    runcolor=cyan
}
\usepackage[pdftex]{color}
\usepackage{braket}
\usepackage{enumerate}
\usepackage[normalem]{ulem}
\usepackage[usenames,dvipsnames]{xcolor}
\usepackage{multirow}
\usepackage{mathtools}
\usepackage{ulem}

\definecolor{orange}{rgb}{1,0.5,0}
\newtheorem{theorem}{Theorem}
\newtheorem{definition}{Definition}
\newtheorem{mylemma}{Lemma}
\newtheorem{proposition}{Proposition}

\newcommand{\ignore}[1]{}

\ignore{
\documentclass[eprintnumbers,amsmath,amssymb,onecolumn,a4paper,caption ]{article}
\usepackage{amsfonts}
\usepackage{amssymb}
\usepackage{mathrsfs}
\usepackage{mathbbold}
\usepackage{bbm}
\usepackage{mathrsfs}
\usepackage{dcolumn}
\usepackage{bm}
\usepackage{times,epsfig,amssymb,amsmath}
\usepackage{float}
\usepackage{subfigure}

\usepackage{color}

}

\begin{document}

\title{Logarithmic light cone, slow entanglement growth, and quantum memory}
\author{Yu~Zeng}\thanks{zengyunk@gmail.com}
\affiliation{Shandong Inspur Intelligence Research Institute Co., Ltd., Jinan 250100, China}
\affiliation{Shandong Yunhai Guochuang Innovative Technology Co., Ltd., Jinan 250101, China}
\author{Alioscia~Hamma}\thanks{alioscia.hamma@unina.it}
\affiliation{Dipartimento di Fisica `Ettore Pancini', Universit\`a degli Studi di Napoli Federico II, Via Cintia 80126, Napoli, Italy}
\affiliation{INFN, Sezione di Napoli, Italy}
\author{Yu-Ran Zhang}
\affiliation{School of Physics and Optoelectronics, South China University of Technology, Guangzhou 510640, China}
\author{Qiang Liu}
\affiliation{Shandong Inspur Intelligence Research Institute Co., Ltd., Jinan 250100, China}
\affiliation{Shandong Yunhai Guochuang Innovative Technology Co., Ltd., Jinan 250101, China}
\author{Rengang Li}
\affiliation{Shandong Inspur Intelligence Research Institute Co., Ltd., Jinan 250100, China}
\affiliation{Shandong Yunhai Guochuang Innovative Technology Co., Ltd., Jinan 250101, China}
\author{Heng~Fan}
\affiliation{School of Physical Sciences, University of Chinese Academy of Sciences, Beijing 100049, China}
\author{Wu-Ming Liu}
\affiliation{School of Physical Sciences, University of Chinese Academy of Sciences, Beijing 100049, China}

\begin{abstract}
Effective light cones, characterized by Lieb-Robinson bounds, emerge in nonrelativistic local quantum systems. Here, we present several analytical results derived from logarithmic light cones (LLCs). Possible origins of LLCs include the one-dimensional (1D) disordered XXZ model and a phenomenological model of many-body localization (MBL). In the LLC regime, we prove that, for arbitrary spatial dimensions and any initial pure state, entanglement growth is upper-bounded by logarithmic time with an additional subleading \emph{double-logarithmic} correction---arising from a real asymptotic solution of the \emph{Lambert W} function---valid up to the asymptotic time limit. In the context of the 1D disordered XXZ model, this result resolves the ambiguity in distinguishing between logarithmic and power-law fits of entanglement growth in numerical studies; we also propose a falsifiable phenomenological functional form for the entanglement growth that agrees with existing numerical results. We show that information scrambling is logarithmically slow in the LLC regime. Furthermore, we demonstrate that the LLC supports long-lived quantum memories---quantum codes with macroscopic code distance and lifetimes that scale exponentially with system size---under unitary time evolution. Our analytical results provide benchmarks for future numerical studies of the MBL regime at large time scales.
\end{abstract}


\maketitle


\section{Introduction}
The speed of light imposes a fundamental constraint on relativistic quantum field theory in Minkowski space, dubbed `microcausality' \cite{Haag1964}. It states that two physical operators separated by spacelike intervals must commute:
\begin{eqnarray}
[O_1(x), O_2(y)]=0 ~~\text{for}~~(x-y)^2<0\nonumber
\end{eqnarray}
implying that a measurement outside the light cone of another operator cannot have any influence on it.
Remarkably, a similar property exists for nonrelativistic quantum lattice models with local interactions, where an effective linear light cone emerges characterized by so-called Lieb-Robinson bounds (LRBs) \cite{liebrobinson0}. Explicitly, for any bounded local operators $O_X$ and $O_Y$ supported on lattice sets $X$ and $Y$ with distance $\operatorname{dist}(X,Y)=l$,
\begin{eqnarray}\label{linearLRB}
\|[O_X(t), O_Y]\|\leq\operatorname{exp}(-\Omega(l)) ~~\text{for}~~v_L\lvert t\lvert-l<0,
\end{eqnarray}
where  the state-independent effective ``speed of light'' $v_L$ is the Lieb-Robinson velocity. This inequality implies that, although the information spreads outside the light cone is not zero, it  vanishes exponentially \cite{liebrobinson2}. In the continuum limit, these bounds become sharp \cite{Cramer&Eisert2008}. These bounds can also be extended to general Markovian dynamics \cite{Poulin2010}, specific classes of infinite-dimensional systems, notably harmonic systems \cite{Cramer&Eisert2008} and systems that are commutator bounded \cite{PS&Hamma2010}, signified by the fact that all that needs to be physically finite is the exchange of energy.

LRBs are at the basis of several fundamental theorems in many-body physics, such as the exponential clustering theorem \cite{Hastings2004a,Hastings2004b,Hastings2006,Nachtergaele2006}, the Lieb-Schultz-Mattis theorem in Higher dimensions \cite{Hastings2004a}, the quantization of Hall conductance \cite{Hastings2015} and the concept of quasi-adiabatic continuation \cite{HastingsWen2005}. However, it was noticed that there could be tighter bounds in localized quantum systems \cite{Burrell2007}. Ref. \cite{Hamza2012} proved that the one-dimensional (1D) disordered XX model with single-particle localization exhibits a bounded light cone. Recent analytical works \cite{Elgart2022,Elgart2023} proved that the 1D disordered XXZ model can exhibit a logarithmic light cone (LLC) for a finite system size.

Evidence of LLCs has been observed over the past decades through the slow dynamics of entanglement. Numerical and experimental simulations for microscopic models \cite{Znidaric2008,Bardarson2012,Deng2017,Yang&Chamon2017,Xu2018,Lukin2019} suggest that entanglement grows \emph{logarithmically} with time from an initial product state in the many-body localization (MBL) \cite{Basko2006,Oganesyan2007,Pal2010,Abanin2019RMP} regime, distinguishing it from integrable and chaotic models, which exhibit ballistic growth \cite{Calabrse2005,Kim2013}, as well as from diffusive systems, which show power-law growth \cite{Rakovszky2019,Znidaric2020}. Formal arguments based on local integrals of motion (LIOMs), or $l$-bit models of MBL systems, provide a transparent explanation of this phenomenon: the slow entanglement growth results from dephasing dynamics of conserved $l$-bits with exponentially decaying interactions  \cite{Serbyn2013a,Serbyn2013b,Huse2014,Chandran2015,KimI2014,Znidaric2018}. However, whether MBL exists as a stable dynamical phase in the thermodynamic limit remains unclear based on current numerical studies \cite{Panda2020}. Due to limited system sizes and accessible time scales, current numerical results cannot unambiguously be extrapolated to the asymptotic limit (see the recent review \cite{Sierant2024} and the references therein). From the perspective of slow entanglement growth, numerical results suggesting logarithmic growth in microscopic models are not conclusive, as Ref. \cite{Sierant2024} indicates that the logarithmic growth of entanglement entropy is not easily distinguishable from power-law growth. Furthermore, MBL is a robust ergodicity-breaking quantum phenomenon that applies to arbitrary physically relevant initial states, in contrast to fine-tuned settings like many-body scars \cite{Serbyn2021}. For these reasons, analytical results on entanglement growth in the MBL regime, valid for arbitrary initial pure states, are still needed.

Recent works \cite{Kiefer2020,Kiefer2021a,Kiefer2021b,Sierant2022,Ghosh2022} have also investigated number entropy (NE) \cite{Lukin2019}, which contributes to the entanglement entropy arising from the particle number fluctuations between subsystems, particularly in particle-number-conserved systems such as the XXZ model. Some studies \cite{Kiefer2020,Kiefer2021a,Kiefer2021b,Sierant2022} found that the NE increases \emph{double-logarithmically} with time, interpreting this as evidence of the absence of MBL phases, since the unbounded growth of NE may contradict predictions from the $l$-bit model. However, these numerical results have been subject to debate \cite{ Kiefer2022comment,Ghosh2022reponse}. More recent numerical work \cite{Chavez2023} detects the double-logarithmic growth of NE in an $l$-bit model of MBL, which suggests that such growth of NE is not sufficient to rule out MBL.

On another note, the rapid advancement of quantum technologies \cite{google2023,Krinner2022,Zhao2022,Bravyi2022,Lucas2023,Jones2020,Levitin2022} promises to realize quantum memories (QMs), ideally self-correcting and analogous to robust classical memories \cite{Brown2016}. Unfortunately, beyond the thermal bath, perturbed unitary evolutions alone can thermalize a QM, making it unreliable \cite{Brown2016,Zeng2016,Kay2009,Pastawski2010,Kay2011}. Accidental local errors after non-equilibrium evolution can become global errors that cannot be corrected. Recent numerical results of out-of-time-order correlators (OTOCs), commonly used to diagnose information scrambling and operator growth, exhibit logarithmically slow scrambling in the MBL regime \cite{Huang2017,Chen2017,ChenY2016,He2017,Fan2017,Swingle2017prb,Nahum2018prb,Sahu-Xu-Swingle2019prl,Xu2019prr,KimSW2021,KimM2023}. These results hint that LLCs may preserve QMs for exponentially long times. 

In this paper, we formally define an LLC for local quantum systems and analytically discuss its physical consequences. Our main result is Theorem \ref{Slogt} and its proof, which states that the LLC implies that entanglement grows at most logarithmically with time, with an additional \emph{double-logarithmic} correction. Specifically, the time-dependent growth of bipartite entanglement entropy is upper bounded by $2|\partial|\xi[(\alpha+1)\ln t+\ln\ln t]$, where $|\partial|$, defined in Theorem \ref{Slogt}, is the area of the subsystem boundary; $\xi$ and $\alpha$ are positive real numbers from Definition \ref{LLCdef}. Existing proofs of logarithmic entanglement growth based on LLCs are limited by the techniques used, which only apply to 1D models and initial product states \cite{KimI2014,Eisert2006,Burrell2007}. In contrast, Theorem \ref{Slogt} and its proof are valid for arbitrary finite spatial dimensions and any initial pure state. The leading term of the bound, $2|\partial|\xi(\alpha+1)\ln t$, confirms the conjecture presented in Ref. \cite{KimI2014} regarding the bound in higher-dimensional systems.

In combination with recent analytical work by Elgart and Klein \cite{Elgart2023}, which goes beyond the $l$-bit assumption, our result demonstrates that the entanglement growth is logarithmically slow for the finite-size 1D disordered XXZ model in the MBL regime. This result is independent of the initial state and is valid to the asymptotic time limit. By comparing recent numerical results on entanglement growth in the 1D disordered XXZ model, we propose a falsifiable phenomenological interpretation involving the double-logarithmic correction, in which the leading term corresponds to the configurational entropy and the subleading term corresponds to the number entropy. We also prove that LLCs support long-lived QMs---quantum error-correcting codes with macroscopic distance and a lifetime that scales exponentially with system size---under unitary time evolution at zero temperature.

\section{LLC and MBL}
Consider a local nonrelativistic quantum system on a $D$-dimensional lattice $\Lambda$, a metric space of sites with a distance function $\operatorname{dist}\left(\cdot, \cdot\right)$. The linear size of $\Lambda$ is $L$ satisfying $\mid\Lambda\mid\sim L^D$. Here, $\mid\cdots\mid$ is the cardinality of a set. The distance between two sets is $\operatorname{dist}\left(X, Y\right)=\operatorname{min}_{i\in X, j\in Y}\operatorname{dist}\left(i, j\right)$, and the diameter of a set is $\operatorname{diam}\left(X\right)=\operatorname{max}_{i, j\in X}\operatorname{dist}\left(i, j\right)$. The Hilbert space $\mathscr{H}=\otimes_{i\in\Lambda}\mathscr{H}_{i}$ is the tensor product of local finite-dimensional Hilbert spaces defined on lattice sites. Without loss of generality, we consider spin-$\frac{1}{2}$ systems where dim$(\mathscr{H}_{i})=2$. The Hamiltonian $H=\sum_{j\in\Lambda}h_j$ is local and bounded, meaning that the support of each $h_j$ has a finite diameter and $\parallel h_j\parallel$ is finite. The unitary time-evolution operator is $U(t)=e^{-iHt}$ and the operator in Heisenberg picture is $O(t)=U^\dagger(t)OU(t)$.
 We define LLC as
\begin{definition}\label{LLCdef}
A quantum system possesses an LLC if, for any local operators $O_X$ and $O_Y$ supported on sets $X$ and $Y$ separated by a distance $l=\operatorname{dist}(X,Y)$, there exist real positive numbers $C$, $\alpha$, and $\xi$ such that the following inequality holds for all $t\in\mathbb{R}$:
\begin{equation}\label{LLC}
\parallel[O_X(t),O_Y]\parallel\leq C\parallel O_X\parallel\parallel O_Y\parallel \lvert t\rvert ^\alpha e^{-\frac{l}{\xi}}.
\end{equation}
\end{definition}

A typical possible scenario inducing the LLC is MBL. Ref. \cite{KimI2014} finds that a phenomenological model of MBL system can lead to an LLC (see also Appendix \ref{AppendixA}). The phenomena of MBL suggest that extensively many LIOMs---dubbed $l$-bits \cite{Serbyn2013b,Huse2014,Serbyn2013b,Huse2014,Ros2015,Imbrie2017}---emerge. Refs. \cite{Imbrie2016PRL, Imbrie2016JSP} prove the existence of the MBL phase in a 1D system under a physically reasonable assumption. Although the existence of stable MBL phases in higher dimensions remains controversial \cite{Chandran2016,DeRoeck2017a,DeRoeck2017b}, certain numerical evidences suggest $l-$bits exists in two-dimensional, even three-dimensional models with random local potential \cite{Thomson2018,Wahl2019,Chertkov2021}. Recent researches argue that MBL phases may be stable in two-dimension at strong enough quasi-periodic modulation \cite{Agrawal2022,Crowley2022,Strkalj2022}. Beyond phenomenological LIOMs, Ref. \cite{Elgart2023} demonstrates that LLC can emerge in a 1D microscopic model of MBL. The authors proved that a variant form of Eq. (\ref{LLC}) can be fulfilled in the 1D disordered XXZ model for finite system sizes, while the prefactor $C$ is polynomially dependent on the system size.
It is worth noting that, MBL is only one possible origin of LLC, and one cannot exclude other origins. We show several physical consequences of LLC in the following sections.

\section{LLC and Slow entanglement growth}
Consider a simply connected region $A$ and its complement, $B=\Lambda-A$, and a bipartite Hilbert space $\mathscr{H}=\mathscr{H}_A\otimes\mathscr{H}_B$. The bipartite entanglement of a pure state $\rho=|\Psi\rangle\langle\Psi|$ is measured by the entanglement entropy \cite{Amico2008}: $S(|\Psi\rangle)=S(\rho)=-\operatorname{Tr}\rho_A\log_2\rho_A$, where $\rho_A=\operatorname{Tr}_B \rho$. It is well known that the entanglement growth, $\Delta S(t)=S(U(t)|\Psi\rangle)-S(|\Psi\rangle)$, for a generic local quantum system is upper bounded by a linear function of time \cite{liebrobinson2}, as a direct consequence of a constant entanglement rate \cite{Acoleyen2013}. In the LLC regime, one cannot simply apply a rate of entanglement growth to achieve a tighter bound. Existing analytical results have shown that,  for 1D space and an initial product state, the entanglement grows at most logarithmically with time \cite{Burrell2007,KimI2014}. In what follows, we prove the main theorem of this paper, which is valid for arbitrary finite spatial dimensions and any initial pure state.
\begin{theorem}\label{Slogt}
Let $H=\sum_{j\in\Lambda}h_j$ be a time-independent Hamiltonian that is local, bounded, and satisfies the condition of LLC in definition \ref{LLCdef}. For an arbitrary initial pure state $|\Psi\rangle$, the entanglement growth $\Delta S(t)=S(U(t)|\Psi\rangle)-S(|\Psi\rangle)$ is bounded by
\begin{eqnarray}\label{DeltaS}
\Delta S\!\leq\!2|\partial|\xi[(\alpha\!+\!1)\ln t\!+\!\ln\!\ln t\!+\!\ln\left((1\!+\!e^{1/\xi})CJ|\partial|\xi\right)\!+\!\mathcal{O}(1)],\nonumber\\
\end{eqnarray}
where $\xi$, $\alpha$, and $C$ are defined as in Definition \ref{LLCdef}; the boundary between subsystems $A$ and $B$ is given by $\partial\equiv\{j\in\Lambda~|h_j$ acts nontrivially on both $A$ and $B$\}, and $J=\operatorname{max}_{n\in\partial}\{\parallel h_n\parallel\}$.
\end{theorem}

\begin{figure}
\centering
\includegraphics[width=0.5\linewidth]{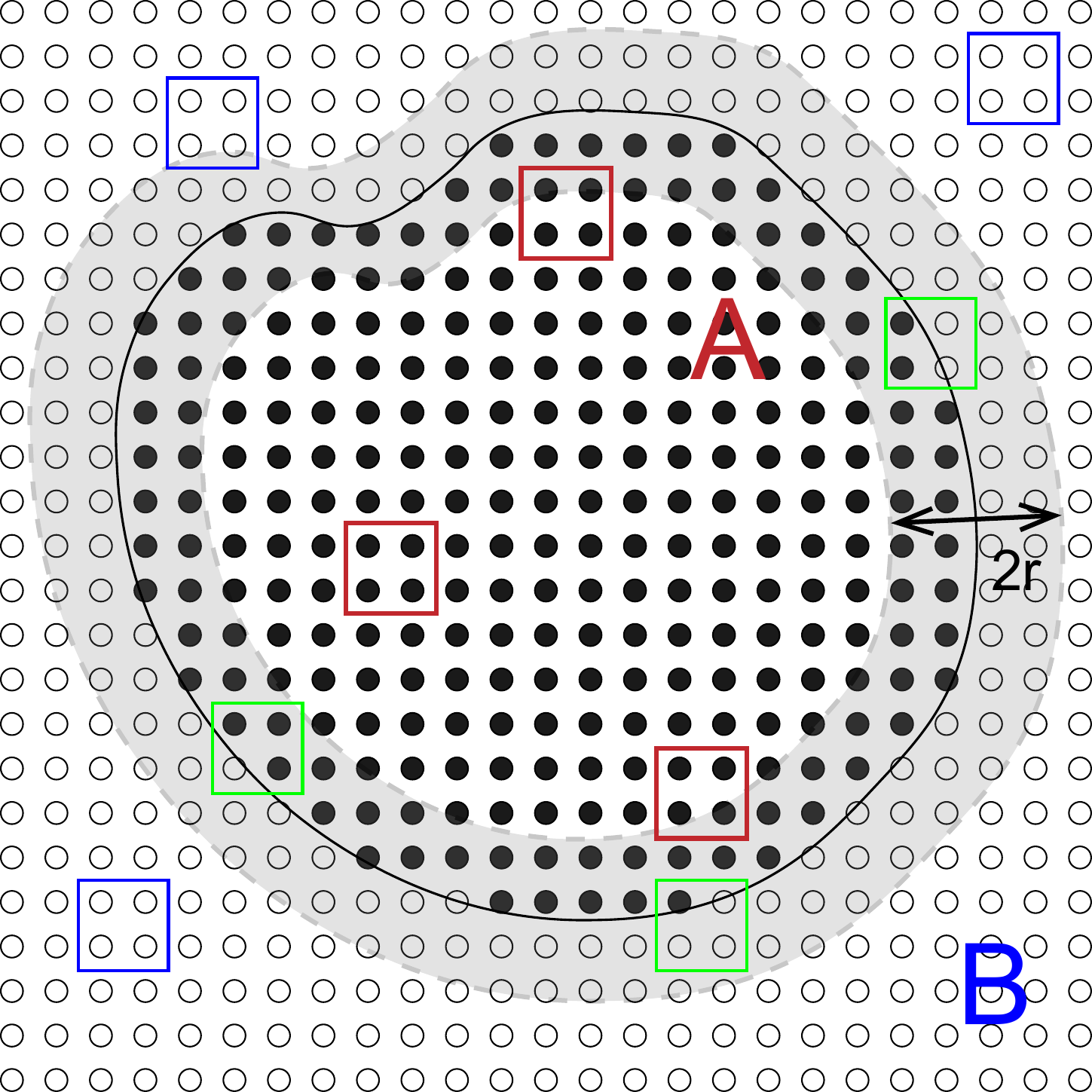}\\
\caption{ Illustration of a bipartite two-dimensional local quantum (spin-$\frac{1}{2}$) system. The Hilbert space is the tensor product of local spin space residing on lattice sites belonging to $A$ (black dots) and $B$ (white dots). The Hamiltonian interactions are contained within compact regions. Examples of the interactions are illustrated by the square box, which are classified into three types: acting on $A$ (red), $B$ (blue), or both $A$ and $B$ (green). The support of $U_{\partial}$ in Eq. (\ref{dUpartial}) is illustrated by the grey ribbon with width $2r$.}\label{bipartite}
\end{figure}

\begin{proof} Being local, the
 Hamiltonian can be partitioned into $H=H_A+H_B+H_{\partial}$, $H_A$ ($H_B$) only acting on $A$ ($B$) nontrivially while the boundary term $H_{\partial}=\sum_{n\in\partial}h_n$ acting on both $A$ and $B$ nontrivially (see Fig. \ref{bipartite}). The unitary time evolution operator is $U(t)=\operatorname{e}^{-it(H_A+H_B+H_{\partial})}$. We decompose $U(t)$ as
\begin{eqnarray}\label{timeevo}
U(t)=U_{1}(t)U_0(t).
\end{eqnarray}
$U_0(t)$ has the tensor product form $U_0(t)=\operatorname{e}^{-itH_A}\bigotimes \operatorname{e}^{-itH_B}$, so denoting $|\Psi_0\rangle\equiv U_0|\Psi\rangle$, we have $S(|\Psi_0\rangle)=S(|\Psi\rangle)$. 
$U_{1}(t)$ satisfies the differential equation
\begin{eqnarray}\label{dU1}
\frac{\partial}{\partial t}U_{1}(t)=-i \sum_{n\in\partial}h_n(t)U_{1}(t),
\end{eqnarray}
where $h_n(t)=U(t)h_nU^\dagger(t)$.

We introduce an auxiliary parameter $r$, and $h_n(t,r)=\frac{1}{\operatorname{Tr}_{S_r}\left(\mathbbm{1}_{S_r}\right)} \operatorname{Tr}_{S_r}\left[h_n(t)\right] \otimes \mathbbm{1}_{S_r}$, where $S_r$ is the set of sites having distance at least $r$ from $n$. Indeed,
\begin{eqnarray}\label{Haar}
h_n(t,r)=\int\operatorname{d}\mu(V_r)V_rh_n(t)V_r^{\dagger},
\end{eqnarray}
where the integral is over unitary operators supported on $S_r$ with Haar measure $d\mu$ \cite{liebrobinson2}. We define $U_{\partial}(t,r)$ satisfying
\begin{eqnarray}\label{dUpartial}
\frac{\partial}{\partial t}U_{\partial}(t,r)&=&-i\sum_{n\in\partial}h_n(t,r)U_{\partial}(t,r).
\end{eqnarray}
The support of $U_{\partial}$ forms a ribbon with length $|\partial|$ and width $2r$, so $U_{\partial}$ nontrivially acts on a Hilbert space with dimension $d_r=2^{2r|\partial|}$ (see Fig. \ref{bipartite}), where we have dropped an order one prefactor in the exponent for succinctness. Further, we define $U_{\delta}(t,r)$ satisfying
\begin{eqnarray}\label{dUdelta}
\frac{\partial}{\partial t}U_{\delta}(t,r)&=&-iU_{\partial}^{-1}\sum_{n\in{\partial}}\left[h_n(t)-h_n(t,r)\right]U_{\partial}U_{\delta}.\nonumber\\
&=&-iH_{\delta}(t,r)U_{\delta}(t,r).
\end{eqnarray}
The effective Hamiltonian $H_{\delta}$ can be rearranged to
\begin{eqnarray}
H_{\delta}(t,r)&\equiv&\sum_{r^\prime=r}^{\infty}U_{\partial}^{-1}\sum_{n\in{\partial}}\left[h_n(t,r^\prime+1)-h_n(t,{r^\prime})\right]U_{\partial}\nonumber\\
&=&\sum_{r^\prime=r}^\infty\tilde{H}_{\delta}(t,r^\prime),
\end{eqnarray}
where the support of each term $\tilde{H}_{\delta}(t,r^\prime)$ forms a ribbon with length $|\partial|$ and width $2(r^\prime+1)$.
Combining Eqs. (\ref{dU1}), (\ref{dUpartial}) and (\ref{dUdelta}), We have $U_1(t)=U_{\partial}(t,r)U_{\delta}(t,r)$.

The following strategy is to apply two conclusions about entanglement growth, namely \emph{small incremental entangling} \cite{Bravyi2007,Acoleyen2013} to bound entanglement growth generated by $U_{\delta}$, and \emph{small total entangling} \cite{Bravyi2007,Bennett2003} to bound entanglement growth generated by $U_{\partial}$. We take two steps separately for clarity.

\emph{\textbf{Step 1.}}
We first bound $\Delta S_1=S\left(U_{\delta}|\Psi_0\rangle\right)-S\left(|\Psi_0\rangle\right)$.
The rate of entanglement growth is
\begin{eqnarray}
\Gamma(t^{\prime},r)&\equiv&\frac{\partial}{\partial t^{\prime}}S(\rho)\nonumber\\
&=&i\operatorname{Tr}\left(H_{\delta}(t^{\prime},r)[\rho,\operatorname{log}\rho_A\otimes\mathbbm{1}_B]\right)\nonumber\\
&=&\sum_{r^\prime=r}^{\infty}i\operatorname{Tr}\left(\tilde{H}_{\delta}(t^{\prime},r^\prime)[\rho,\operatorname{log}\rho_A\otimes\mathbbm{1}_B]\right),\nonumber
\end{eqnarray}
where $\rho=U_{\delta}(t^{\prime},r)|\Psi_0\rangle\langle\Psi_0|U_{\delta}(t^{\prime},r)^{-1}$.
\emph{Small incremental entangling} \cite{Acoleyen2013} states that
\begin{eqnarray}\label{EIrate}
\Gamma(t^{\prime},r)\leq c_1\sum_{r^\prime=r}^\infty\parallel\tilde{H}_{\delta}(t^{\prime},r^\prime)\parallel\operatorname{log_2}(d_{r^\prime+1}),
\end{eqnarray}
where $c_1=\mathcal{O}(1)$, and $d_{r^\prime+1}=2^{|\partial|(r^\prime+1)}$. $\parallel\tilde{H}_{\delta}(t^{\prime},r^\prime)\parallel$ can be bounded by the LLC condition in definition \ref{LLCdef}, and triangle inequality:
\begin{eqnarray}\label{triangleinequality}
\parallel\!\!\tilde{H}_{\delta}(t^{\prime}\!,\!r^\prime)\!\!\parallel\!&\leq&\!\sum_{n\in\partial}\!\parallel\!\! h_n(t^{\prime}\!,\!r^\prime\!+\!\!1)\!-\!h_n(t^{\prime})\!\!\parallel\!\!+\!\!\parallel\!\!h_n(t^{\prime})\!-\!h_n(t^{\prime}\!,\!r^\prime)\!\!\parallel\nonumber\\
&\leq&(1+e^{1/\xi})CJ|\partial|t^{\prime\alpha}\operatorname{exp}\left(-\frac{r^\prime+1}{\xi}\right).
\end{eqnarray}
Here we use Eq. (\ref{Haar}) that gives $\parallel h_n(t^{\prime})-h_n(t^{\prime},r^\prime)\parallel\leq\int d\mu(V_{r^\prime})\parallel[V_{r^\prime},h_n(t^{\prime})]\parallel\leq CJt^{\prime\alpha}e^{-r^\prime/\xi}$. So, Eq. (\ref{EIrate}) can be bounded as
\begin{eqnarray}
\Gamma(t^{\prime},r)&\leq&(1+e^{1/\xi})c_{1} CJ|\partial|^2t^{\prime\alpha}\sum_{r^\prime=r}^\infty(r^\prime+1)\operatorname{exp}\left(-\frac{r^\prime+1}{\xi}\right)\nonumber\\
&\leq&(1+e^{1/\xi})c_{1}CJ|\partial|^2t^{\prime\alpha}\int_{r}^\infty dr^\prime(r^\prime)\operatorname{exp}\left(-\frac{r^\prime}{\xi}\right)\nonumber\\
&=&(1+e^{1/\xi})c_{1}CJ|\partial|^2t^{\prime\alpha}(\xi r+\xi^2)e^{-r/\xi}\nonumber
\end{eqnarray}
Therefore, the entanglement growth generated by $U_{\delta}$ can be bounded by
\begin{eqnarray}
\Delta S_1=\int_{0}^{t}dt^{\prime}\Gamma(t^{\prime},r)\leq C_1|\partial|^2\frac{t^{\alpha+1}}{\alpha+1}(\xi r+\xi^2)e^{-r/\xi},\nonumber
\end{eqnarray}
where we combine all the constants into $C_1=(1+e^{1/\xi})c_{1}CJ$.

\emph{\textbf{Step 2.}} We move on to bound $\Delta S_2=S\left(U_{\partial}U_{\delta}|\Psi_0\rangle\right)-S\left(U_{\delta}|\Psi_0\rangle\right)$. Noting that $U_{\partial}$ has support with cardinality $\log_2(d_{r})$, \emph{small total entangling} guarantees that $\Delta S_2\leq2\times1/2\log_2(d_{r})=2|\partial|r$ \cite{Bennett2003} (see also Appendix \ref{AppendixB}).

After these two steps, we have
\begin{eqnarray}\label{EI}
\Delta S&=&\Delta S_1+\Delta S_2\nonumber\\
&\leq&2|\partial|r+C_1|\partial|^2\frac{t^{\alpha+1}}{\alpha+1}(\xi r+\xi^2)e^{-r/\xi}.
\end{eqnarray}
We denote the R.H.S of Eq. (\ref{EI}) as $f(r,t)$, and we can get the optimal bound by selecting the minima of all $t$ slices. Explicitly, for a fixed $t$, 
\begin{eqnarray}\label{LambertW}
\frac{\partial f(r,t)}{\partial r}=2|\partial|-C_1|\partial|^2\frac{t^{\alpha+1}}{\alpha+1}re^{-r/\xi}=0,
\end{eqnarray}
and
\begin{eqnarray}\label{2nd-derivative}
\frac{\partial^2 f(r,t)}{\partial r^2}=C_1|\partial|^2\frac{t^{\alpha+1}}{\alpha+1}\left(\frac{r}{\xi}-1\right)e^{-r/\xi}>0.
\end{eqnarray}
\begin{figure}
\centering
\includegraphics[width=0.5\linewidth]{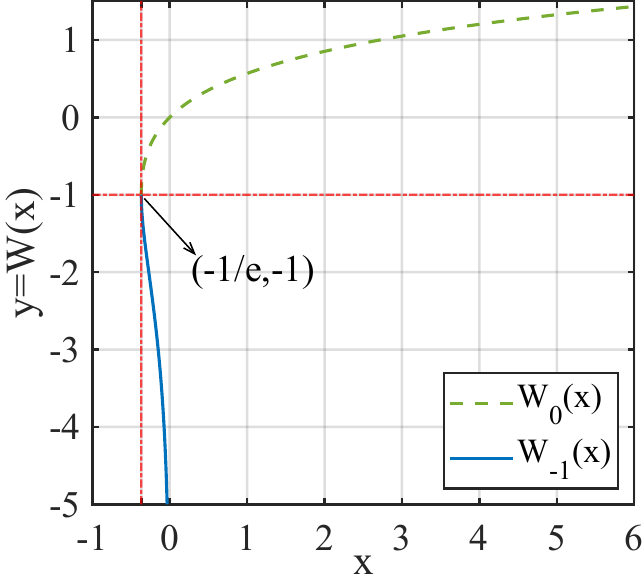}\\
\caption{ Illustration of two real branches of \emph{Lambert W function} $y=W(x)$ where $x=ye^y$. Green dashed line: $W_0(x)$; blue solid line: $W_{-1}(x)$.  $W_{-1}(x)$ is the legitimate solution in Eq. (\ref{ysolution}).}\label{lambertW}
\end{figure}

Eq. (\ref{LambertW}) has the form of \emph{Lambert W function}: $ye^y=x$, where $x=-\frac{2(\alpha+1)}{C_1|\partial|\xi t^{\alpha+1}}$ and $y=-r/\xi$. It can be solved for real number $y$ only if $x\geq-\frac{1}{e}$ (see Fig. \ref{lambertW}) . $y$ has two solutions if $-\frac{1}{e}\leq x<0$, which are $y=W_0(x)\geq-1$ and $y=W_{-1}(x)\leq-1$. Nevertheless, Eq. (\ref{2nd-derivative}) selects $y=W_{-1}(x)$ to be the legitimate solution which has the asymptotic form \cite{LambertW}:
\begin{eqnarray}\label{ysolution}
y=W_{-1}(x)=\ln(-x)-\ln(-\ln(-x))-o(1).
\end{eqnarray}
From Eqs. (\ref{EI}) and (\ref{LambertW}), the optimal bound is $\Delta S\leq2|\partial|\xi(-y+1-1/y)$. Combining Eq. (\ref{ysolution}), we have (see Appendix \ref{AppendixC})
\begin{eqnarray}\label{DeltaS2}
\Delta S\!\leq\!2|\partial|\xi[(\alpha\!+\!1)\ln t\!+\!\ln\ln t\!+\!\ln(\frac{e}{2}C_1|\partial|\xi)\!+\!o(1)].
\end{eqnarray}
Eq. (\ref{DeltaS}) is obtained by inserting $C_1=(1+e^{1/\xi})c_{1}CJ$.
\end{proof}

In Theorem \ref{Slogt}, the leading term in Eq. (\ref{DeltaS}) indicates that the entanglement between the bipartite subsystems grows at most as $2|\partial|\xi(\alpha+1)\ln t$, where $|\partial|$ is the area of the subsystem boundary, $\alpha$ is the temporal growth exponent, and $\xi$ is the characteristic length of the LLC in Definition \ref{LLCdef}. This confirms the conjecture in Ref. \cite{KimI2014} for arbitrary finite $D$ spatial dimensions. Recall that, for a generic local quantum system, the entanglement entropy increases at most linearly in time ($\sim|\partial|t$), satisfying the area law. In contrast, within the LLC regime, the entanglement growth is at most logarithmic in time and still obeys the area law. More precisely, the entanglement growth is proportional to the volume of a shell encasing the boundary---$|\partial|$ multiplied by the thickness $2\xi$. In two-dimensional space, for instance, this shell becomes a ribbon of width $2\xi$ (see the grey ribbon in Fig. \ref{bipartite}). Considering that the system and subsystem volumes scale as $L^{D}$, the boundary area satisfies $|\partial|\sim L^{D-1}$, and the initial pure state obeys an entanglement area law, Theorem \ref{Slogt} implies that the entanglement entropy requires at least exponentially long times to saturate to a volume law. This is consistent with the numerical result of entanglement growth in a 1D $l$-bit model \cite{Chavez2023} over sufficiently long times.

Theorem \ref{Slogt} also applies to a typical 1D microscopic model of MBL: the disordered XXZ model. It is well-known that the logarithmic entanglement growth is consistent with the $l$-bit model of MBL \cite{Abanin2019RMP,Serbyn2013a,Huse2014}. However, as noted in Ref. \cite{Sierant2024}, current numerical simulations of the disordered XXZ model cannot reliably distinguish between power-law and logarithmic growth due to the limited accessible time scales and system sizes. Consequently, the asymptotic behavior of entanglement growth remains uncertain. Elgart and Klein \cite{Elgart2023} recently proved that there exists a non-trivial parameter region in the finite-size disordered XXZ model where an LLC emerges. Combining this result with Theorem \ref{Slogt}, we demonstrate that the entanglement entropy grows at most logarithmically with time in this MBL regime of the disordered XXZ model. This conclusion is valid for arbitrary initial pure states, not just product states, and holds throughout the time domain up to the asymptotic limit  $t\rightarrow e^{\Theta(L)}$. Note that the MBL phase is defined as a robust dynamical phase of matter characterized by ergodicity breaking for arbitrary initial state, and necessarily concerns the asymptotic limit $t\rightarrow\infty$ and $L\rightarrow\infty$. We stress that, due to the limitation of numerical method, the existence and stability of the MBL phase is not clear \cite{Panda2020,Sierant2024}. From the perspective of slow entanglement growth, Ref. \cite{Sierant2024} argues that the ambiguity of distinguishing asymptotic logarithmic growth from a power-law growth hinders clear conclusions about the existence and the extent of the MBL phase. However, our result suggests that this MBL regime can be extended to the asymptotic time limit, $t\rightarrow e^{\Theta(L)}$, for arbitrary initial pure states. Unfortunately, achieving the thermodynamic limit $L\rightarrow\infty$ remains unfeasible at the current stage.

\section{Phenomenological interpretation involving the double-logarithmic correction}
Now, we discuss the peculiar double-logarithmic correction, $2|\partial|\xi\ln\ln t$, in Eq. (\ref{DeltaS}). At first glance, this subleading term is meaningless in an inequality, as the leading logarithmic term can dominate it. Essentially, the upper bound on entanglement growth in Eq. (\ref{DeltaS}) is derived from the LLC condition in Definition \ref{LLCdef}. The parameters in the LLC condition, which may be derived from a particular microscopic model, are generally not optimal. Consequently, the subleading correction to the bound is not meaningful, since the prefactor of the leading term is generally not exact. However, by comparing with existing numerical results, we argue that a phenomenological interpretation involving the double-logarithmic correction may still be meaningful.

The $U(1)$ symmetry of the 1D XXZ model ensures that the entanglement entropy can be split into configurational entropy and NE \cite{Lukin2019}. Numerical studies \cite{Kiefer2020,Kiefer2021a,Kiefer2021b,Sierant2022} investigating the entanglement growth from an initial product state---which is an eigenstate of the total $z$ spin operator---indicate that the entanglement growth can be well fitted by a functional form: $\Delta S_{XXZ}(t)=\mu\ln t+\nu\ln\ln t+\operatorname{const.}$. The leading logarithmic term corresponds to configurational entropy, while the subleading double-logarithmic term corresponds to NE.
Returning to Eq. (\ref{DeltaS}), phenomenologically, we conjecture that the entanglement growth within LLC has the functional form $\Delta S_{phe}(t)=2|\partial|\tilde{\xi}(\tilde{\alpha}+1)\ln t+2|\partial|\tilde{\xi}\ln\ln t+\operatorname{const.}$, where the positive parameters $\tilde{\xi}$ and $\tilde{\alpha}$ are not derived from any microscopic models but serve as adjustable variables used to match the coefficients from `experimental' data fitting, such that $2|\partial|\tilde{\xi}(\tilde{\alpha}+1)=\mu$ and $2|\partial|\tilde{\xi}=\nu$. For the 1D disordered XXZ model, the area of the entanglement-cut boundary, $|\partial|$,  is an order-one constant.

In fact, the phenomenological conjecture $\Delta S_{phe}(t)$ is falsifiable. Since $\tilde{\alpha}$ is positive, the above substitution imposes an additional constraint on the coefficients in the numerical data fitting: if $\Delta S_{phe}(t)$ agrees with $\Delta S_{XXZ}(t)$, then it must hold that $\frac{\mu}{\nu}-1=\tilde{\alpha}>0$. Surprisingly, to the best of our knowledge, existing numerical results, including those from the disordered XXZ model \cite{Kiefer2021a,Kiefer2021b,Sierant2022} and a random circuit $l$-bit model \cite{Chavez2023}, satisfy this constraint \cite{alpha}. Another support for $\Delta S_{phe}(t)$ is that Refs. \cite{Kiefer2021a,Kiefer2021b} show that the coefficients satisfy $\mu\sim\nu\sim1/W^3$ where $W$ is the disorder strength of the magnetic field in the XXZ model. Although based on small length and time scales, these numerical results are compatible with $\Delta S_{phe}(t)$ when noting that $\mu\sim\nu\sim\tilde{\xi}$. According to Definition \ref{LLCdef}, we can interpret $\tilde{\xi}$ as the characteristic length of the disordered XXZ model. It is reasonable to expect that $\tilde{\xi}$  inversely correlates with $W$, and the numerical data fitting suggests $\tilde{\xi}\sim1/W^3$. Furthermore, an explicit LLC profile for the disordered XXZ model can be obtained through the data-fitting coefficients of configurational entropy and NE growth. By referring to Definition \ref{LLCdef}, we suggest that the contours $l/\tilde{\xi}-\tilde{\alpha}\operatorname{log}t=\theta$, where $\theta$ is a positive real variable, depict the light-cone profile.


\section{Slow information scrambling and long-lived quantum memory}
For spin operators $W_{r}$ and $V_{r^\prime}$ and density matrix $\rho_\beta=e^{-\beta H}/\operatorname{tr}\left(e^{-\beta H}\right)$, the OTOC is defined by $F_{\beta}(r,t)=\operatorname{tr}\left(\rho_\beta W^\dagger_{0}(t)V^\dagger_{r} W_{0}(t)V_{r}\right)$ which is closely related to the expectation value of squared commutator $C_{\beta}(r,t)=\frac{1}{2}\operatorname{tr}\left(\rho_\beta\left[W_{0}(t), V_{r}\right]^\dagger\left[W_{0}(t), V_{r}\right]\right)=1-\operatorname{Re}\left[F_{\beta}(r,t)\right]$. Especially, $F_{0}(r,t)$ is real at infinite temperature. The analytical or numerical results of the contours $C_{0}(r_{\theta},t)=\theta$ suggest that the light-cone profile of some typical local quantum models are linear with broadened wave front \cite{Khemani2018,Xu&Swingle2019,Xu&Swingle2020,Xu&Swingle2022}, where $r_{\theta}=v_B t+\left(\frac{t^p}{\lambda}\operatorname{ln}\frac{1}{\theta}\right)^{\frac{1}{1+p}}=\mathcal{O}(t)$ with a nonnegative constant $p$ . Here $v_B$ is the ``butterfly'' velocity \cite{Xu&Swingle2022,Roberts2016} and it is compatible with the linear LRB in Eq. (\ref{linearLRB}). In the LLC regime (see Definition \ref{LLCdef}),  $C_{\beta}(r,t)\leq\frac{1}{2}\parallel\left[W_{0}(t), V_{r}\right]\parallel^2\leq C^\prime e^{2(\alpha\xi\operatorname{ln}t-r)/\xi}$ implies that $v_B=0$ \cite{Sahu-Xu-Swingle2019prl,Xu2019prr}; the quantum information scrambling and operator growth are logarithmically slow, which are observed in MBL systems \cite{Huang2017,Chen2017,ChenY2016,He2017,Fan2017,Swingle2017prb,Nahum2018prb,Sahu-Xu-Swingle2019prl,Xu2019prr,KimSW2021,KimM2023}.

We consider a QM whose code space is the degenerate ground space of a local Hamiltonian $H_0$ with topological order (TO) \cite{kitaev:2003}. There exists $L^\ast=\Omega(L)$, such that for any operator $O$ with diameter $\leq L^\ast$, the ground-space projector $P_0$ satisfies $P_0 O P_0\propto P_0$ \cite{Bravyi2010prl}. This topological QM \cite{dennis} has a macroscopic code distance according to the \emph{quantum error-correction conditions} \cite{NielsenChuang,Gottesman2009}: the error space, denoted by $\mathcal{E}$, can be corrected, iff $P_0E^\dagger EP_0\propto P_0$ for all $E\in\mathcal{E}$. The topological QM is expected to be self-correcting as the TO is robust against local perturbations \cite{Bravyi2010a}: the gap of  $H_s=H_0+sV$ is stable for small local perturbations $sV$, and there exist a scalar $z$ and $\epsilon=e^{\left(-\Omega(L)\right)}$, such that the ground-space projector $P_s$ satisfies $\parallel P_s OP_s-zP_s\parallel\leq\epsilon$. We say that $P_s$ has TO to accuracy $(L^\ast, \epsilon)$ \cite{liebrobinson2,Hastings2011}. However, the encoded initial state will typically not be the eigenstate of the perturbed Hamiltonian \cite{Brown2016}. An accidental local errors can spread to be non-local after a time of $\mathcal{O}(L)$ and become uncorrectable. Equivalently, the code space $U(t)P_{0}U^\dagger(t)$ can tolerate errors with only $\mathcal{O}(1)$ diameter at the same time scale. The following proposition shows that the LLC can support long-lived topological QMs over unitary time evolution. This proposition is a direct generalization of the argument in the regime of linear light cone \cite{liebrobinson2}.

\begin{proposition}
Suppose that the initial code space $P$ has TO to accuracy $(L^\ast,\epsilon)$, and the system possesses an LLC. Then, there exists a $t^\ast=\operatorname{exp}\left(\Omega(L)\right)$, such that for any $t\leq t^\ast$, $P(t)=U(t)PU^\dagger(t)$ is topologically ordered to accuracy $(L^\ast/2,\epsilon^\prime)$, where $\epsilon^\prime=\operatorname{exp}\left(-\Omega(L)\right)$.
\end{proposition}
\begin{proof}
For any operator $O$ supported on a set with diameter smaller than $L^\ast/2$, and setting $\parallel O\parallel=1$ without loss of generality, we introduce $\tilde{O}(t)=\int\operatorname{d}\mu(V_{L^\ast/4})V_{L^\ast/4}O(t)V_{L^\ast/4}^{\dagger}$. The support of unitary $V_{L^\ast/4}$ has distance at least $L^\ast/4$ from the support of $O$ and $d\mu(V_{L^\ast/4})$ is the Haar measure. The diameter of the support of $\tilde O(t)$ is less than $L^\ast$ \cite{Zeng2025}, so there exists a scalar $z$ such that $\parallel P\tilde O(t)P-zP\parallel\leq\epsilon$.
Using the triangle equality, we have $\parallel PO(t)P-zP\parallel\leq\parallel PO(t)P-P\tilde O(t)P\parallel+\parallel P\tilde O(t)P-zP\parallel$.
Applying the technique used in Eq. (\ref{triangleinequality}), the first term is bounded as $\parallel PO(t)P-P\tilde O(t)P\parallel\leq Ce^{\frac{-L^\ast/4+\alpha\xi\ln{t}}{\xi}}$, where $C$ depends at most algebraically on $L$. We take $t^\ast=\exp\left(\frac{L^\ast}{8\alpha\xi}\right)$. For $t\leq t^\ast$, we have $\parallel PO(t)P-zP\parallel\leq\epsilon^\prime$, where $\epsilon^\prime=\epsilon+Ce^{\frac{-L^\ast}{8\xi}}$ is exponentially small in $L$.
\end{proof}

\section{Conclusions}
We formally define an LLC for local nonrelativistic quantum systems and discuss its physical consequences. We prove that the entanglement growth is upper-bounded by logarithmic time with a double-logarithmic correction in the LLC regime for arbitrary finite spatial dimensions and any initial pure state. In the context of the 1D disordered XXZ model,  combining the analytical result of Ref. \cite{Elgart2023}, we conclude that the entanglement grows at most logarithmically with time up to the asymptotic time limit. This result resolves the ambiguity in distinguishing between logarithmic and power-law growth of entanglement entropy in numerical studies \cite{Sierant2024}. Evaluating the MBL phase---defined as a robust, ergodicity-breaking dynamical phase of matter---necessarily requires taking the asymptotic limit $L,t\rightarrow\infty$ and considering arbitrary physically relevant initial states. Due to the limited length and time scales accessible in current numerical studies, the existence of an MBL phase remains unclear. Nevertheless, from the perspective of slow entanglement growth, our theoretical analysis extends this MBL regime to the asymptotic time limit $t\rightarrow e^{\Theta(L)}$ for arbitrary initial pure states,  although the thermodynamic limit remains inaccessible.

By comparing current numerical results \cite{Kiefer2021a,Kiefer2021b,Sierant2022,Chavez2023}, where the entanglement growth is fitted by the functional form $\Delta S_{XXZ}(t)=\mu\ln t+\nu\ln\ln t+\operatorname{const.}$, with Eq. (\ref{DeltaS}), we propose a phenomenological conjecture: $\Delta S_{phe}(t)=2|\partial|\tilde{\xi}(\tilde{\alpha}+1)\ln t+2|\partial|\tilde{\xi}\ln\ln t+\operatorname{const.}$, which matches $\Delta S_{XXZ}(t)$ by setting $\mu=2|\partial|\tilde{\xi}(\tilde{\alpha}+1)$ and $\nu=2|\partial|\tilde{\xi}$. This conjecture is falsifiable, as it implies $\mu>\nu$, and all existing numerical data, to the best of our knowledge, satisfy this condition \cite{alpha}. This observation is nontrivial, since there is no \emph{a priori} reason that the prefactor of the logarithmic term (configurational entropy) must exceed that of the double-logarithmic term (number entropy). Given its testability, we suggest that future numerical studies investigate the validity of $\Delta S_{phe}(t)$ over larger system sizes, longer timescales, and broader parameter ranges. Furthermore, an explicit LLC profile can be numerically extracted from the fitted coefficients $\mu$ and $\nu$.

Finally, by generalizing the well-known argument from the regime of linear light cone \cite{liebrobinson2}, we show that the LLC supports long-lived topological QMs under perturbed unitary evolution: although the encoded information will eventually become irretrievable even at zero temperature, the lifetime of a topological QM can scale exponentially with system size. Gapped topologically ordered phases form the foundation of topological (or self-correcting) QMs, and it is well known that quantum spin systems exhibiting such order can exist only in two or higher spatial dimensions \cite{Chen2011}. Although the existence of MBL phases remains unsettled, if mechanisms---such as MBL---could realize LLCs in two or three dimensions, topological QMs would likely exhibit dynamical robustness. The folk wisdom that `MBL retains the memory of initial state' may find a practical realization in this setting. Effective light cones provide versatile tools for studying dynamical properties in quantum information and topological quantum phases \cite{liebrobinson2,Zeng2025}. We also anticipate that future analytical works may extend nonlinear light-cone structures to quasiperiodic systems, as has been achieved for disordered spin chains \cite{Elgart2022,Elgart2023}.



\begin{acknowledgments}
Y. Z. thanks D. Toniolo for useful communication.
A. H. is supported in part by PNRR MUR project PE0000023-NQSTI and
PNRR MUR project CN $00000013$. Y. R. Z. is supported in part by the National Natural Science Foundation of China (Grant No.~12475017), the Natural Science Foundation of Guangdong Province (Grant No.~2024A1515010398), and the Startup Grant of South China University of Technology (Grant No.~20240061). H. F. is supported in part by NSFC (grant Nos. 92265207, T2121001). W. -M. L. is supported in part by National Key R\&D Program of China (grant Nos.  2021YFA1400900, 2021YFA0718300, 2021YFA1402100) and NSFC (grant Nos. 61835013, 12174461, 12234012).
\end{acknowledgments}


\section{APPENDIX}
\subsection{Logarithmic light cone from a phenomenological model of MBL}\label{AppendixA}
In this section, we derive that the logarithmic light cone, with $\alpha=1$ in Eq. (\ref{LLC}) presented in the main text, can emerge from a phenomenological model of Many-body localization (MBL) \cite{Chandran2015}. Other positive $\alpha\neq 1$ may emerge from generic phenomenological models.  The derivation follows Kim, Chandran, and Abanin \cite{KimI2014}, and we only fill in more details.

We assume the existence of the MBL phase in $D$ dimensions, and the Hamiltonian is the sum of quasi-local integral of motions (LIOMs): $H=\sum_{j\in\Lambda}\tilde{h}_j$, where $\left[H, \tilde{h}_j\right]=0$ and $\left[\tilde{h}_i, \tilde{h}_j\right]=0$ for all $i, j\in\Lambda$ \cite{KimI2014,Chandran2015}. The norm of the Hamiltonian is extensive, namely $\parallel H\parallel\sim|\Lambda|$, so $\parallel\tilde{h}\parallel=\mathcal{O}(1)$. $\tilde{h}$ is quasi-local, meaning that for any lattice site $i$ and operator $O_X$,
\begin{eqnarray}\label{quasilocal}
\parallel\left[\tilde{h}_j, O_X\right]\parallel\leq\operatorname{const.}\parallel O_X\parallel e^{-\operatorname{dist}(i, X)/\xi^\prime}.
\end{eqnarray}

We denote the commutator as $f(t)=\left[O_{X}(t),O_{Y}\right]$, where $O_{X}(t)=e^{iHt}O_{X}e^{-iHt}$. Here, the subscript $X$ denotes the support of the operator $O_{X}$, and $Y$ denotes the support of $O_{Y}$. The distance between $X$ and $Y$ is $\operatorname{dist}(X, Y)=l$. We define the $X(l/2)\subset\Lambda$ the set of sites at a distance no larger than $l/2$ from $X$, and we denote the complement of $X(l/2)$ by $X^c(l/2)$. For the $D$-dimensional lattice $\Lambda$, the cardinality of $X(l/2)$ is $|X(l/2)|=\mathcal{O}(l^D)$. The Hamiltonian can be decomposed as
\begin{eqnarray}
H=\tilde{H}_{X(l/2)}+\tilde{H}_{X^c(l/2)}.
\end{eqnarray}
where
\begin{eqnarray}
\tilde{H}_{X(l/2)}=\sum_{j\in X(l/2)}\tilde{h}_{j}.
\end{eqnarray}
and
\begin{eqnarray}
\tilde{H}_{X^c(l/2)}=\sum_{j\in X^c(l/2)}\tilde{h}_{j}.
\end{eqnarray}

To bound $\parallel\left[O_X(t),O_{Y}\right]\parallel$, we apply the time derivative
\begin{widetext}
\begin{eqnarray}\label{dfdt}
f^\prime(t)&=&i\left[\left[H, O_{X}(t)\right], O_{Y}\right]\nonumber\\
&=&i\left[\left[\tilde{H}_{X(l/2)}, O_{X}(t)\right], O_{Y}\right]+i\left[\left[\tilde{H}_{X^c(l/2)}, O_{X}(t)\right], O_{Y}\right]\nonumber\\
&=&-i\left[\left[O_X(t), O_Y\right], \tilde{H}_{X(l/2)}\right]-i\left[\left[O_{Y}, \tilde{H}_{X(l/2)}\right], O_{X}(t)\right]+i\left[\left[\tilde{H}_{X^c(l/2)}, O_X(t)\right], O_{Y}\right].
\end{eqnarray}
We pick out the second and third terms:
\begin{eqnarray}
\delta(t)=-i\left[\left[O_{Y}, \tilde{H}_{X(l/2)}\right], O_{X}(t)\right]+i\left[\left[\tilde{H}_{X^c(l/2)}. O_{X}(t)\right], O_{Y}\right],
\end{eqnarray}
The norm of $\delta(t)$ is bounded by
\begin{eqnarray}
||\delta(t)||\leq2\parallel\left[O_{Y}, \tilde{H}_{X(l/2)}\right]\parallel\parallel O_{X}\parallel+2\parallel\left[O_{X}, \tilde{H}_{X^c(l/2)}\right]\parallel\parallel O_{Y}\parallel,
\end{eqnarray}
where we use $\left[H, \tilde{H}_{X^c(l/2)}\right]=0$.
\end{widetext}
By combining Eq. (\ref{quasilocal}), we can bound the norm of the commutator as follows:
\begin{eqnarray}
\parallel\left[O_{Y}, \tilde{H}_{X(l/2)}\right]\parallel&\leq&\sum_{j\in X(l/2)}\parallel\left[\tilde{h}_{j}, O_{Y}\right]\parallel\nonumber\\
&\leq&\mathcal{O}(l^D)\parallel O_{Y}\parallel e^{-l/2\xi^\prime}\nonumber\\
&=&\parallel O_{Y}\parallel\mathcal{O}(e^{-l/2\xi^\prime}),
\end{eqnarray}
and
\begin{eqnarray}
\parallel\left[O_{X}, \tilde{H}_{X^c(l/2)}\right]\parallel&\leq&\sum_{j\in X^c(l/2)}\parallel\left[\tilde{h}_{j}, O_{X}\right]\parallel\nonumber\\
&\leq&\operatorname{const.}\parallel O_{X}\parallel\sum_{r=l/2}^{\infty}r^{D-1}e^{\left(-r/\xi^\prime\right)}\nonumber\\
&\leq&\operatorname{const^\prime.}\parallel O_{X}\parallel\int_{l/2}^{\infty}dr r^{D-1}e^{\left(-r/\xi^\prime\right)}\nonumber\\
&=&\parallel O_{X}\parallel\mathcal{O}(e^{-l/2\xi^\prime}).
\end{eqnarray}
Thus we get
\begin{eqnarray}\label{deltanorm}
||\delta(t)||\leq\parallel O_{X}\parallel\parallel O_{Y}\parallel\mathcal{O}\left(e^{-l/2\xi^\prime}\right).
\end{eqnarray}

The time derivative in Eq. (\ref{dfdt}) can be written in the discrete form for $\epsilon\rightarrow 0$,
\begin{eqnarray}
f(t+\epsilon)&=&f(t)-i\left[f(t),\tilde{H}_{X(l/2)}\right]\epsilon+\delta(t)\epsilon\nonumber\\
&=&e^{i\tilde{H}_{X(l/2)}\epsilon}f(t)e^{-i\tilde{H}_{X(l/2)}\epsilon}+\delta(t)\epsilon+o(\epsilon^2).
\end{eqnarray}

So we have
\begin{eqnarray}
\parallel f(t+\epsilon)\parallel-\parallel f(t)\parallel\leq\parallel\delta(t)\parallel\epsilon+o(\epsilon^2).
\end{eqnarray}
Given that $f(0)=0$, we can obtain
\begin{eqnarray}
\parallel f(t)\parallel\leq\int_{0}^{t}dt^\prime\parallel\delta(t^\prime)\parallel.
\end{eqnarray}
Combining Eq. (\ref{deltanorm}), we have
\begin{eqnarray}
\parallel\left[O_{X}(t),O_{Y}\right]\parallel\leq t\parallel O_{X}\parallel\parallel O_{Y}\parallel\mathcal{O}\left(e^{-l/2\xi^\prime}\right).
\end{eqnarray}

\subsection{A proof of a lemma: small total entangling}\label{AppendixB}
In this section, we provide a proof of \emph{small total entangling} \cite{Bravyi2007}, which was introduced as Lemma 1 in Ref. \cite{Bennett2003}. Our goal is to adapt the proof from Ref. \cite{Bennett2003} to make it more readily understandable for readers from diverse disciplines. The notations follow the main text. $A\subset\Lambda$ is a subset of lattice sites and its complement is $B=\Lambda-A$. The Hilbert space is the tensor product of spin-half space defined on the lattice sites, and it is bipartite: $\mathscr{H}=\mathscr{H}_A\otimes\mathscr{H}_B$.
\begin{mylemma}[Small total entangling]
$U_{AB}=U_{ab}\otimes\mathbbm{1}_{\overline{a}}\otimes\mathbbm{1}_{\overline{b}}$ is a unitary operator acting on $\mathscr{H}$, where $a\subset A$ ($b\subset B$) and $\overline{a}=A-a$ ($\overline{b}=B-b$). Denote $d=\operatorname{min}\{2^{a},2^{b}\}$. Then, for arbitrary state $\rho_{AB}$, the entanglement between $A$ and $B$ satisfies:
\begin{eqnarray}
E\left(U_{AB}\rho_{AB}U^\dagger_{AB}\right)-E\left(\rho_{AB}\right)\leq2\log_2 d.
\end{eqnarray}
Here, $E$ is an entanglement measure \cite{Plenio2007} satisfying the property of additivity, meaning $E(\sigma\otimes\rho)=E(\sigma)+E(\rho)$, $\forall\sigma$ and $\rho$. In particular, if $\rho_{AB}=|\Psi\rangle_{AB}$ is a pure state, the entanglement is measured by the entanglement entropy:
\begin{eqnarray}
S\left(U_{AB}|\Psi\rangle_{AB}\right)-S\left(|\Psi\rangle_{AB}\right)\leq2\log_2 d.
\end{eqnarray}
\end{mylemma}

\begin{proof}
Without loss of generality, we consider $a\leq b$.

Suppose Alice holds the subsystem $A$ and Bob holds the subsystem $B$. We introduce the maximally entangled state
\begin{eqnarray}
|\Phi_d\rangle_{\alpha\beta}=\frac{1}{\sqrt{d}}\sum_{j=1}^{d}|j\rangle_\alpha|j\rangle_\beta,
\end{eqnarray}
where $E(|\Phi_d\rangle_{\alpha\beta})=S(|\Phi_d\rangle_{\alpha\beta})=\log_2 d$. We need to consume a maximally entangled state to perform teleportation \cite{Bennett1993}, which is LOCC (Local Operations and Classical Communication) that dose not increase the entanglement \cite{Plenio2007}.

\emph{\textbf{Step 1.}}
Let Alice and Bob share a $|\Phi_d\rangle_{\alpha\beta}$, and then teleport the qubits in $a$ from Alice to Bob. Bob now get extra qubits labeled by $\beta$, and Alice loses her qubits in $a$.

\emph{\textbf{Step 2.}}
Bob perform $U$ on his qubits in $\beta$ and $b$. This operation is a local operation that does not affect the entanglement.

\emph{\textbf{Step 3.}}
Let Alice and Bob share a $|\Phi_d\rangle_{\alpha^\prime\beta^\prime}$, and then teleport the qubits in $\beta$ from Bob to Alice. Alice now get extra qubits labeled by $\alpha^\prime$, and Bob loses his extra qubits in $\beta$.

\emph{\textbf{Step 4.}}
Change lable $\alpha^\prime$ to $a$.

After these four steps, the initial state $\rho_{AB}$ is changed to $U_{AB}\rho_{AB}U^\dagger_{AB}$. Since the whole progress is LOCC, which does not increase the entanglement, we have
\begin{eqnarray}
E\left(\rho_{AB}\otimes|\Phi_d\rangle_{\alpha\beta}\otimes|\Phi_d\rangle_{\alpha^\prime\beta^\prime}\right)\geq E\left(U_{AB}\rho_{AB}U^\dagger_{AB}\right).
\end{eqnarray}
Applying properties of additivity and maximally entanglement, we get
\begin{eqnarray}
E\left(U_{AB}\rho_{AB}U^\dagger_{AB}\right)-E\left(\rho_{AB}\right)\leq2\log_2 d.
\end{eqnarray}
\end{proof}

\subsection{Derivation of Eq. (\ref{DeltaS2}) in the main text}\label{AppendixC}
The \emph{Lambert W function} reads $ye^y=x$, where $x=-\frac{2(\alpha+1)}{C_1|\partial|\xi t^{\alpha+1}}$ and $y=-r/\xi$. According to the analysis in the main text, the solution is
\begin{eqnarray}\label{solution}
y=W_{-1}(x)=\ln(-x)-\ln(-\ln(-x))-o(1).
\end{eqnarray}
Introduce $\gamma=\left(\frac{C_1 |\partial|\xi}{2(\alpha+1)}\right)^{\frac{1}{\alpha+1}}$ such that $x=-\left(\gamma t\right)^{-(\alpha+1)}$. The first term in Eq. (\ref{solution}) is
\begin{eqnarray}\label{firstterm}
-\ln(-x)&=&(\alpha+1)\ln{t}+(\alpha+1)\ln{\gamma}.
\end{eqnarray}
The second term in Eq. (\ref{solution}) is
\begin{eqnarray}\label{secondterm}
\ln(-\ln(-x))&=&\ln(\alpha+1)+\ln(\ln{\gamma}+\ln{t})\nonumber\\
&=&\ln(\alpha+1)+\ln\ln{t}+\ln\left(1+\frac{\ln{\gamma}}{\ln{t}}\right),
\end{eqnarray}
where $\ln\left(1+\frac{\ln{\gamma}}{\ln{t}}\right)=o(1)$ since $\lim_{t\to\infty}\ln\left(1+\frac{\ln{\gamma}}{\ln{t}}\right)=0$.

According to the analysis in the main text, the optimal bound of entanglement growth is
\begin{eqnarray}
\Delta S\leq2|\partial|\xi(-y+1-1/y),
\end{eqnarray}
where $-1/y=o(1)$. Combining the Eqs. (\ref{solution}), (\ref{firstterm}), and (\ref{secondterm}), we finally get
\begin{widetext}
\begin{eqnarray}
\Delta S&\leq&2|\partial|\xi[(\alpha+1)\ln t+\ln\ln t+\ln(\frac{1}{2}C_1|\partial|\xi)+1+\ln\left(1+\frac{\ln{\gamma}}{\ln{t}}\right)+o(1)]\nonumber\\
&=&2|\partial|\xi[(\alpha+1)\ln t+\ln\ln t+\ln(\frac{e}{2}C_1|\partial|\xi)+o(1)].
\end{eqnarray}
\end{widetext}

%

\end{document}